\documentclass[epjCONF,columns]{svjour} 
\usepackage{graphics}
\usepackage[varg]{txfonts} 
\usepackage[latin1]{inputenc}
\usepackage{epsfig,graphicx}
\usepackage[square,sort,comma,numbers]{natbib}

\def\simlt{\lower.5ex\hbox{$\; \buildrel < \over \sim \;$}}
\def\simgt{\lower.5ex\hbox{$\; \buildrel > \over \sim \;$}}
\newcommand {\PN}{{\rm PN}}
\session-title{Conference Title, to be filled}
\begin{document}
\title{Fake plunges are very eccentric real EMRIs in disguise}
\subtitle{... they dominate the rates and are blissfully ignorant of angular momentum barriers}

\author{Pau Amaro-Seoane\inst{1}\fnmsep\thanks{\email{Pau.Amaro-Seoane@aei.mpg.de}}
        \and
        Carlos F. Sopuerta\inst{2}\fnmsep\thanks{\email{sopuerta@ieec.uab.es}} 
        \and
        Patrick Brem\inst{1}\fnmsep\thanks{\email{Patrick.Brem@aei.mpg.de}}
}
\institute{Max-Planck Institut for Gravitational Physics
(Albert Einstein-Institut) 
\and
Institut de Ci{\`e}ncies de l'Espai (CSIC-IEEC), Campus UAB,
Torre C5 parells, 08193 Bellaterra, Spain  
}
\abstract{
The capture of a compact object in a galactic nucleus by a massive black hole
(MBH) is the best way to map space and time around it.  Compact objects such as stellar black holes on a
capture orbit with a very high eccentricity have been wrongly assumed to be
lost for the system after an intense burst of radiation, which has been
described as a ``direct plunge''. We prove that these very eccentric capture
orbits spend actually a similar number of cycles in a LISA-like detector as
those with lower eccentricities if the central MBH is spinning. Although the
rates are higher for high-eccentricity EMRIs, the spin also enhances the rates
of lower-eccentricity EMRIs. This last kind have received more attention
because of the fact that high-eccentricity EMRIs were thought to be direct plunges and thus negligible. On the other hand,
recent work on stellar dynamics has demonstrated that there seems to be a
complot in phase space acting on these lower-eccentricity captures, since their rates
decrease significantly by the presence of a blockade in the rate at which
orbital angular momenta change takes place. This so-called ``Schwarzschild
barrier'' is a result of the impact of relativistic precession on to the
stellar potential torques, and thus it affects the enhancement on
lower-eccentricity EMRIs that one would expect from resonant relaxation. We
confirm and quantify the existence of this barrier using a statitical sample of
2,500 direct-summation $N-$body simulations using both a post-Newtonian but
also, and for the first time, a geodesic approximation for the relativistic
orbits.  The existence of the barrier prevents ``traditional EMRIs'' from
approaching the central MBH, but if the central MBH is spinning the rate will
be anyway dominated by highly-eccentric extreme-mass ratio inspirals, which
insolently ignore the presence of the barrier, because they are driven by
two-body relaxation.
} 
\maketitle
\section{Introduction}
\label{intro}

One of the most exciting results of modern astronomy is the discovery,
mostly through high-resolution observations of the kinematics of stars and gas,
that most, if not all, nearby bright galaxies harbor a dark, massive, compact
object at their centers. \citep{FF04,Kormendy04}. The most spectacular case is
our own galaxy, the Milky Way.  By tracking and interpreting the stellar dynamics at the centre of our galaxy,
we have the most well-established evidence for the existence of a massive
black hole (MBH).
The close examination of the Keplerian orbits of the
so-called ``S-stars'' (also called S0-stars, where the letter S stands simply
for source) has revealed the nature of the central dark object located at the
Galactic Center. By following one of them, S2 (S02), the mass of SgrA$^*$ was
estimated to be about $3.7\times 10^6\,M_{\odot}$ within a volume with radius
no larger than 6.25 light-hours \citep{SchoedelEtAl03,GhezEtAl03b}.  More
recent data based on 16 years of observations set the mass of the central MBH
to $\sim 4 \times 10^{6} \, M_{\odot}$
\citep{EisenhauerEtAl05,GhezEtAl05,GhezEtAl08,GillessenEtAl09}.  
Observations
of other galaxies indicate that the masses of MBH can reach
a few billion solar masses. The existence of such a MBH population in
the present-day universe is strongly supported by So{\l}tan's argument
that the average mass density of these MBHs agrees with expectations from
integrated luminosity of quasars \citep{Soltan82,YT02}.

To interact closely with the central MBH, stars have to find themselves on
``loss-cone'' orbits, which are orbits elongated enough to have a very close-in
pericenter \citep{FR76,LS77,AS01}. The rate of tidal disruptions can be
established (semi-)analytically if the phase space distribution of stars around
the MBH is known \citep{MT99,SU99,WM04} for estimates in models of observed nearby
nuclei. To account for the complex influence of mass
segregation, collisions and the evolution of the nucleus over billions of
years, detailed numerical simulations are required, however
\citep{DDC87a,DDC87b,MCD91,FB02b,BME04b,FAK06a,KhalEtAl07,PretoAmaroSeoane10,Amaro-SeoanePreto11}.

Many correlations linking the MBH's mass and overall properties of its host
spheroid (bulge or elliptical galaxy) have been discovered. The tightest are
with the spheroid mass \citep{HR04}, its velocity dispersion ($M-\sigma$
relation, \cite{TremaineEtAl02}) and degree of concentration \citep{EGC04}.
Consequently, understanding the origin and evolution of these MBHs necessitates
their study in the context of their surrounding stellar systems.

These observations are difficult, specially for low-mass MBH (ranging
between $10^5$ and $10^7\,M_{\odot}$) and more so for the even less-massive
intermediate MBHs (IMBHs). Nowadays using adaptive optics we could optimistically hope to get a
handful of measurements of stellar velocities of such targets about $\sim 5$ kpc
away in some ten years.  Nevertheless, we need a bright reference star to guide
us; the requirement of an astrometric reference system is crucial.

Therefore, in order to follow the same scheme of detection of inner stellar
kinematics and of the masses of MBHs in distant galaxies, we need the Very
Large Telescope interferometer and one of the next-generation instruments,
either the VSI or GRAVITY \citep{GillessenEtAl06,EisenhauerEtAl08}. Only then can we
improve our astrometric accuracy by the necessary factor of $\sim 10$. This could allow us
to follow the motion of stars orbiting a central MBH in a galaxy with the
resolution we need to extract the mass of the black hole from the kinematics
of the stars.

This is what makes an experiment such as eLISA
\citep{Amaro-SeoaneEtAl2012,Amaro-SeoaneEtAl2012b} so appealing from the
viewpoint of an astrophysicist, since it can be envisaged as a magnifying glass
that will look much deeper and with more detail in the areas of interest.
However, this is only true for compact stars. Whilst main-sequence stars are
tidally disrupted when approaching the central MBH, compact objects (stellar
black holes, neutron stars, and white dwarfs) slowly spiral into the MBH and
are swallowed after some $\sim 10^5$ orbits in the eLISA band. At the
closest approach to the MBH, the system emits a burst of GWs which contains
information about the spacetime geometry, the masses of the system, and the
spins of the MBH. We can regard each such burst as a snapshot of the system.  
This is what makes EMRIs
so appealing: a set of $\sim 10^5$ bursts of GWs radiated by {\em one} system
will tell us with the utmost accuracy about the system itself, it will test
general relativity \citep{Sopuerta2010}, it will tell us about the distribution of dark objects in
galactic nuclei and globular clusters and, thus, we will have a new
understanding of the physics of the process.  Besides, new phenomena, unknown
and unanticipated, are likely to be discovered.

On the other hand, this process is also directly linked to the  growth of MBHs:
Although there is an emerging consensus about the growth of large-mass MBHs
thanks to So{\l}tan's argument, MBHs with masses up to $10^7\,M_{\odot}$, such
as our own MBH in the Galactic Centre (with a mass of $\sim
4\times10^6\,M_{\odot}$), are enigmatic. There are many different arguments to
explain their masses: accretion of multiple stars from arbitrary directions
\cite{Hills75}, mergers of compact objects such as stellar-mass black holes and
neutron stars \cite{QS90} or IMBHs falling on to the MBH
\citep{PortegiesZwartEtAl06}, or by more peculiar means such as accretion of
dark matter \cite{Ostriker00} or collapse of supermassive stars
\citep{Hara78,ST79,Rees84,PauTesi04,Begelman10}. Low-mass MBHs and, thus, the
early growth of {\em all} MBHs, remain a conundrum.

eLISA is in this regard also attractive, for it will scrutinize exactly the
range of masses fundamental to the understanding of the origin and growth of
supermassive black holes.  By extracting the information encoded in the GWs of
this scenario, we can determine the mass of the central MBH with a ridiculous
relative precision of $\sim 10^{-4}$.  Additionally, the mass of the compact
object which falls into the MBH and the eccentricity of the orbit will be
recovered from the gravitational radiation with a fractional accuracy of also
$\sim 10^{-4}$. All this means that eLISA will not be ``just'' the ultimate test
of general relativity, but an exquisite probe of the spins and range of masses
of interest for theoretical and observational astrophysics and cosmology.

\section{Spinning MBHs}

If the central MBH has a mass larger than $10^7\,M_{\odot}$, then the signal of
an inspiraling stellar black hole, even in its last stable orbit (LSO) will
have a frequency too low for detection. On the other hand, if it is
less massive than $10^4\,M_{\odot}$, the signal will also be quite weak unless
the source is very close. This is why one usually assumes that the mass range
of MBHs of interest in the search of EMRIs for eLISA is between
$[10^7,\,10^4]\,M_{\odot}$. Nonetheless, if the MBH is rotating rapidly, then
even if it has a mass larger than $10^7\,M_{\odot}$, the LSO will be closer to
the MBH and thus, even at a higher frequency, the system should be detectable.
This would push the total mass to a few $\sim 10^7\,M_{\odot}$.

For a binary of an MBH and a stellar black hole to be in the eLISA band, it has
to have a frequency of between roughly $10^{-4}$ and $1$ Hz. The emission of GWs
is more efficient as they approach the LSO, so
that eLISA will detect the sources when they are close to the LSO line. The
total mass required to observe systems with frequencies between $10^{-4}$ and 
$1$ Hz and is of $10^4 - 10^7\,M_{\odot}$. For masses larger than
$10^7\,M_{\odot}$ the frequencies close to the LSO will be too low, so
that their detection will be very difficult. On the other hand, for a total
mass of less than $10^3\,M_{\odot}$ we could in principal detect them at an
early stage, but then the amplitude of the GWs would be rather low.

As the star spirals
down towards the MBH, it has many opportunities to be deflected back by
two-body encounters onto a ``safer orbit'' \citep{AH03,Amaro-SeoaneEtAl07},
hence even the definition of a loss cone is not straightforward. Once again,
the problem is compounded by the effects of mass segregation and resonant
relaxation, to mention two main complications. As a result, considerable
uncertainties are attached to the (semi-)analytical predictions of capture
rates and orbital parameters of EMRIs.

Naively one could assume that the inspiral time is dominated by GW emission and
that if this is shorter than a Hubble time, the compact object will become an
EMRI.  This is wrong, because one has to take into account the relaxation of
the stellar system.  Whilst it certainly can increase the eccentricity of the
compact object, it can also perturb the orbit and circularize it, so that the
required time to inspiral in, $t^{}_{\rm GW}$, becomes larger than a Hubble time.
The condition for the small compact object to be an EMRI is that it is on an orbit
for which $t^{}_{\rm GW} \ll (1-e)\,t^{}_{\rm r}$ \citep{Amaro-SeoaneEtAl07}, with
$t^{}_{\rm r}$ the {\em local} relaxation time. When the binary has a
semi-major axis for which the condition is not fulfilled, the small compact
object will have to be already on a so-called ``plunging orbit'', with $e\ge
e_{\rm plunge} \equiv 1-4\,R^{}_{\rm Schw}/a$, where $R^{}_{\rm Schw}$ is the
Schwarzschild radius of the MBH, i.e. $R^{}_{\rm Schw} =
2GM^{}_{\bullet}/c^{2}$, with $M^{}_{\bullet}$ being the MBH mass.  It has been
claimed a number of times by different authors that this would result in a too
short burst of gravitational radiation which could only be detected if it was
originated in our own Galactic Center \citep{HopmanFreitagLarson07} because one needs a
coherent integration of some few thousands repeated passages through the
periapsis in the eLISA bandwidth.

Therefore, such ``plunging'' objects would then be lost for the GW signal,
since they would be plunging ``directly'' through the horizon of the MBH and
only a final burst of GWs would be emitted, and such burst would be (i) very
difficult to recover, since the very short signal would be buried in a sea of
instrumental and confusion noise and (ii) the information contained in the
signal would be practically nil.  There has been some work on the detectability
of such bursts~\citep{RubboEtAl2006,HopmanFreitagLarson07,YunesEtAl2008}, but they would
only be detectable in our galaxy or in the close
neighborhood, but the event rates are rather low, even in the most optimistic
scenarios.

\begin{figure*}
\resizebox{\hsize}{!}
          {\includegraphics[scale=1,clip]{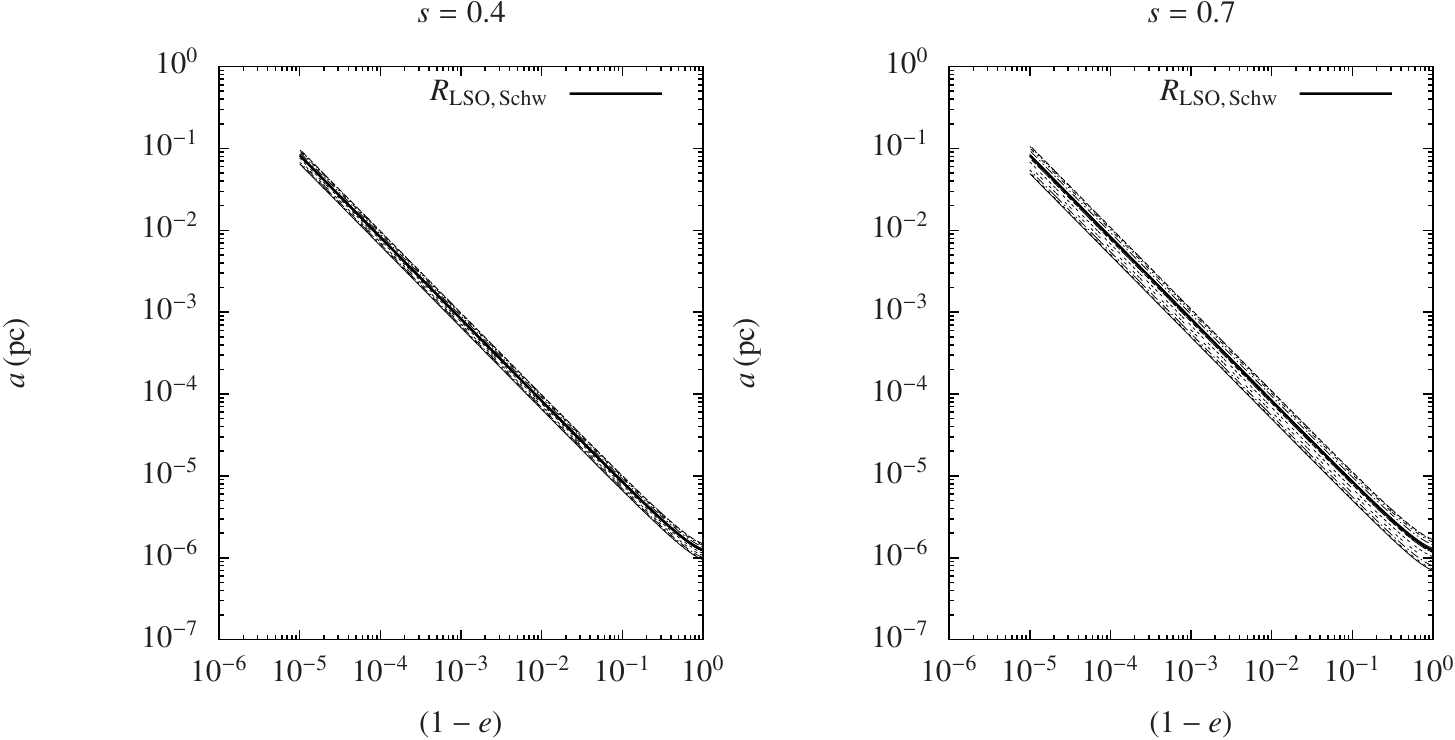}}
\caption
{
LSO for a MBH of mass $4\times10^4\,M^{}_{\odot}$ and a SBH of mass
$m_{\bullet}=10\,M^{}_{\odot}$ for a Kerr MBH of spin $s=0.4$ (left) and $s=0.7$
(right). The Schwarzschild separatrix is given as a solid black line. Curves
above it correspond to retrograde orbits and inclinations of
$\iota=5.72,\,22.91,\,40.10,\,57.29,\,74.48$ and $89.95^{\circ}$ starting from the
last value ($89.95^{\circ}$). In the left panel we can barely see any difference
from the different inclinations due to the low value of the spin.
   }
\label{fig.LSO_Spin0p4_0p7}
\end{figure*}

\begin{figure*}
\resizebox{\hsize}{!}
          {\includegraphics[scale=1,clip]{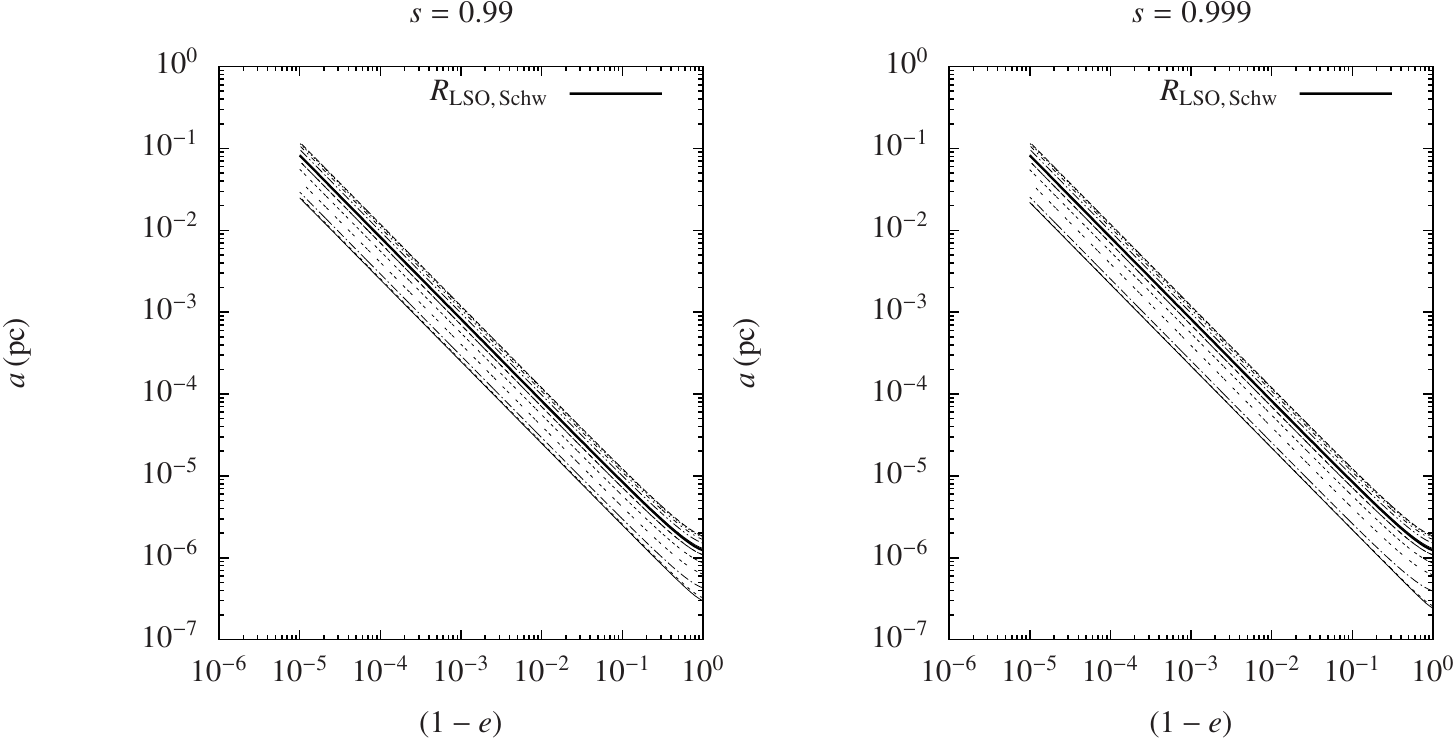}}
\caption
   {
As in figure \ref{fig.LSO_Spin0p4_0p7} but for a spin of $s=0.99$ (left) and
$s=0.999$ (right panel). The larger the spin, the ``further away'' the Kerr LSO
gets from the Schwarzschild LSO.
   }
\label{fig.LSO_Spin0p99_0p999}
\end{figure*}

In figures~\ref{fig.LSO_Spin0p4_0p7} and~\ref{fig.LSO_Spin0p99_0p999} we show
plots of the location of the LSO in the plane $a$ (pc) - $(1-e)$, including the
Schwarzschild separatrix between stable and unstable orbits, $p -6 - 2e = 0$,
for both prograde and retrograde orbits and for different values of the
inclination $\iota$.  Each plot corresponds to a different value of the spin,
showing how increasing the spin makes a difference in shifting the location of
the separatrix between stable and unstable orbits, pushing prograde orbits near
$GM^{}_{\bullet}/c^{2}$ while retrograde orbits are pushed out towards
$9GM^{}_{\bullet}/c^{2}$.  The procedure we have used to build these plots is a
standard one.  Briefly, given a value of the dimensionless spin parameter
$s\equiv a^{}_{\bullet}c^{2}/(GM^{}_{\bullet})$ and a value of the eccentricity
$e$ and inclination angle $\iota$, and we have followed the algorithm,
definitions and notation introduced in \cite{Amaro-SeoaneSopuertaFreitag2012}.

In \cite{Amaro-SeoaneSopuertaFreitag2012} it was estimated the number of cycles that
certain EMRI orbital configurations that were thought to be plunging orbits (or
orbits with no sufficient cycles) in the case of non-spinning MBHs can spend in
a frequency regime of $f\in [10^{-4},1]$ Hz during their last year(s) of
inspiral before plunging into the MBH.  This is important to assess how many of
these EMRIs will have sufficient Signal-to-Noise Ratio (SNR) to be detectable.
It was  found that (prograde) EMRIs that are in a ``plunge'' orbit actually spend  a
significant number of cycles, more than sufficient to be detectable with good
SNR.  The number of cycles has been associated with $N^{}_{\varphi}$ (the
number of times that the azimuthal angle $\varphi$ advances $2\pi$) which is
usual for binary systems.  However, as we have discussed above the structure of
the waveforms from EMRIs is quite rich since they contain harmonics of three
different frequencies.  Therefore the waveforms have cycles associated with the
three frequencies $(f^{}_{r},f^{}_{\theta},f^{}_{\varphi})$ which makes them
quite complex and in principle this is good for detectability (assuming we have
the correct waveform templates). Moreover, these cycles happen just before
plunge and take place in the strong field region very near the MBH horizon.
Then, these cycles should contribute more to the SNR than cycles taking place
farther away from the MBH horizon.

We also estimate the impact on the event rates. Since ``direct plunges''
are actually diguesed EMRIs, although with a higher eccentricity. We prove that

\begin{eqnarray}
\frac{{{a}_{\rm EMRI}^{\rm Kerr}}}{{{a}_{\rm EMRI}^{\rm Schw}}} & = & {\cal W}^{\frac{-5}{6-2\gamma}}(\iota,\,s)\\
\frac{{\dot{N}_{\rm EMRI}^{\rm Kerr}}}{{\dot{N}_{\rm EMRI}^{\rm Schw}}} & = & {\cal W}^{\frac{20\gamma-45}{12-4\gamma}} (\iota,\,s) \,.
\label{eq.NAEMRIW}
\end{eqnarray}

\noindent
In the expression ${\cal W}$ is a function that depends on $\iota$, the
inclination of the EMRI and $s$, its spin\footnote{For the derivation and some
examples of values for ${\cal W}$, we refer the reader to the work of
\cite{Amaro-SeoaneSopuertaFreitag2012}.}.  We also have assumed that the SBHs
distribute around the central MBH following a power-law cusp of exponent
$\gamma$, i.e. that the density profile follows $\rho \propto r^{-\gamma}$
within the region where the gravity of the MBH dominates the gravity of the
stars, with $\gamma$ ranging between 1.75 and 2 for the heavy stellar
components
\citep{Peebles72,BW76,BW77,ASEtAl04,PretoMerrittSpurzem04,AlexanderHopman09,PretoAmaroSeoane10,Amaro-SeoanePreto11}
and see \cite{Gurevich64} for an interesting first idea of this
concept\footnote{The authors obtained a similar solution for how electrons
distribute around a positively charged Coulomb centre.}.

For instance, for a spin of $s=0.999$ and an inclination of $\iota = 0.4\,$rad,
we estimate that ${\cal W}\sim 0.26$ and, thus, $\dot{N}_{\rm EMRI}^{\rm Kerr}
\sim 30$. I.e. {\em we boost the event rates by a factor of 30} in comparison
to a non-rotating MBH.

\section{Analysing a blockade in angular momentum for low-eccentricity EMRIs
with a direct-summation $N-$body sample of 2,500 simulations}

In a gravitational potential with a high degree of symmetry, a test star will
receive gravitational tugs from the rest of the field stars which are not
totally arbitrary and hence do not add up in a random walk way, but
\emph{coherently}.  The potential will prevent stellar orbits from evolving in
an erratic way. In a two-body Keplerian system, a stellar-mass black hole
will orbit around the MBH in a fixed ellipse.  The stellar BH will not feel
random gravitational tugs. It evolves coherently as the result of the action of
the gravitational potential.  When an EMRI approaches the periapsis of its
orbit, we can envisage the situation as a pure two-body problem; initially
Newtonian but later GR effects must be taken into account as the periapsis
grows smaller and smaller.  Nonetheless, as the stellar BH goes back to the
apoapsis, it will feel the surrounding stellar system, distributed in the shape
of a cusp which grows in mass the further away we are from the periapsis.  The
time spent in the region in which we can regard this as a two-body problem is
much shorter than the time in which the stellar BH will feel the rest of the
stellar system. This is particularly true for the kind of objects of our
interest, since the very high eccentricity implies a large semi-major axis. The
time spent on periapsis is negligible as compared with the time spent on
apoapsis.  In that region, the stellar BH feels the graininess of the
potential. The gravitational tugs from other stars will alter its orbit.  The
mean free path in $J$-space of that test stellar BH is very large and thus, it
has a \emph{fast} random walk. Both the magnitude and direction of $J$ of the
stellar black hole are altered.  When the modulus is changed but not the
direction, we talk of ``scalar'' resonant relaxation, and correspondingly when
the direction is changed but not the modulus, ``vector'' resonant relaxation.   

In particular, in the potential of a point mass, orbits are frozen fixed
ellipses that exert a continuous torque on the test star. A test star does not
feel random kicks from all directions.  When we add up the individual
contributions coming from all the rest of the stars on the test star, there is
a residual, non-negligible torque that will influence its evolution. The mean
free path of the star in $J$ space is very large. We will refer to this
phenomenon as \emph{scalar} resonant (or coherent) relaxation, because it can
change both the magnitude of $J$ and the inclination of the orbital plane of
the test star. In this scenario it is possible to alter an initially very
circular orbit and modify it in such a way that the test star will get very
close to the MBH after the torques have acted. i.e., we open a new window for
stars to fall into a capture orbit that will lead to an EMRI.

The impact of coherent relaxation on the production of EMRIs is important.
While the underlying physics of the process is very robust, it is a rather
difficult task to assign values to the different parameters on which the
process depends. A possible way of assigning these was suggested in
\cite{HopmanAlexander06,EilonEtAl09}.  In Figure~6 of \cite{EilonEtAl09}, they
show the rate of EMRIs and plunges in a system in which we take into account
both orthodox or regular relaxation and coherent relaxation, normalised to what
one can expect when only taking into account normal relaxation.  The numerical
simulations of \cite{EilonEtAl09} show that coherent relaxation can enhance the
EMRI rate by a factor of a few over the rates predicted assuming only slow
stochastic two-body relaxation.

Recently, \cite{MerrittEtAl11} estimated with a few direct-summation $N-$body
simulations expanded with a statistical Monte-Carlo study that the production
of ``traditional EMRIs'' via resonant relaxation is markedly decreased by the
presence of a blockade in the rate at which orbital angular momenta change
takes place. This so-called ``Schwarzschild barrier'' is a result of the impact
of relativistic precession on to the stellar potential torques.  Although the
authors find that some particles can penetrate the barrier, EMRIs are
significantly supressed in this scenario. 

We have felt motivated by these results and recently decided to study the
effect with a set of {\em some 2,500} direct-summation $N-$body simulations,
including post-Newtonian corrections and also, for the first time, the
implementation of a solver of geodesic equations in the same code
\citep{BremAmaroSeoaneSopuerta2012}.  For this, we use a modification of the
publicly available {\tt planet} code by Sverre Aarseth, a direct summation
$N$-Body integrator ~\citep{Aarseth99,Aarseth03}. For our study we use several
different methods to account for the general relativistic corrections ot the
Newtonian accelerations:

\begin{itemize}
  \item purely Newtonian
  \item post-Newtonian (PN) corrections
  \item geodesic equations
\end{itemize}

In the purely Newtonian case, the integration is done without modifications to
the acceleration equations. In the PN case we add PN corrections in the
following way:

\begin{equation}
{F}  = \overbrace{{F}_0}^{\rm Newt.}
+\overbrace{\underbrace{c^{-2}{F}_2}_{1\PN} +
\underbrace{c^{-4}{F}_4}_{2\PN}}^{\rm periapsis~shift} +
\overbrace{\underbrace{c^{-5}{F}_5}_{2.5\PN}}^{\rm energy~loss} +
\overbrace{\mathcal{O}(c^{-6})}^{\rm neglected},
\label{eq.F_PN}
\end{equation}

\noindent
i.e., as described in the pioneering work of \cite{KupiEtAl06}.
The individual $F_i$ can be found in \citep{BlanchetFaye01}, their equation (7.16).

Given the high mass ratio for EMRIs, their motion around a MBH can also be
approximated by solving the geodesic equations.  These equations give the exact
trajectory of a test mass particle around a Schwarzschild BH. Unlike the PN
approximation, the geodesic equations are valid even in the last few $r_g$
during a plunge or inspiral, however only in the limit $m_\star/M_\bullet
\rightarrow 0$. Some orbits are expected to migrate towards plunge or inspiral
orbits at pericenter distances of $r_p < 15 \,r_g$, where the errors of the PN
approximation can already be quite significant \citep{YunesBerti2008}. In order
to test the existence of the Schwarzschild barrier at small distances, we
implemented these corrections in the {\tt planet} code.

Although we confirm the blockade, 
we note that the systems the two groups study are representative but
artificial, because they integrate only 50 particles with the same mass
distributed around the MBH. In this regard, the existence of this quenching
must be still explored in more detail, in particular with direct-summation
simulations with realistic number of stars and mass ranges.

In any case, in this scenario the barrier can roughly be fitted in $a$ -- $e$
space as

\begin{equation}
a_{\rm SB} \approx C_{\rm SB} \cdot \left( 1 -e ^2\right)^{-1/3}.
\label{eq.aSB}
\end{equation}

\noindent In the last equation $a_{\rm SB}$ is the semi-major axis in mpc below
which penetration of particles is severely supressed and $C_{\rm SB}$ is a
constant of order unity. 
\cite{MerrittEtAl11} choose $C_{\rm SB} = 0.7$.
We depict the barrier in figure
\ref{fig.Timescales}. As we can see, and as described in the work of
\cite{MerrittEtAl11}, the barrier poses a real problem for stars with small
semi-major axis, and below it, the evolution is dominated by two-body
relaxation.

\begin{figure}
\resizebox{\hsize}{!}
          {\includegraphics[scale=1,clip]{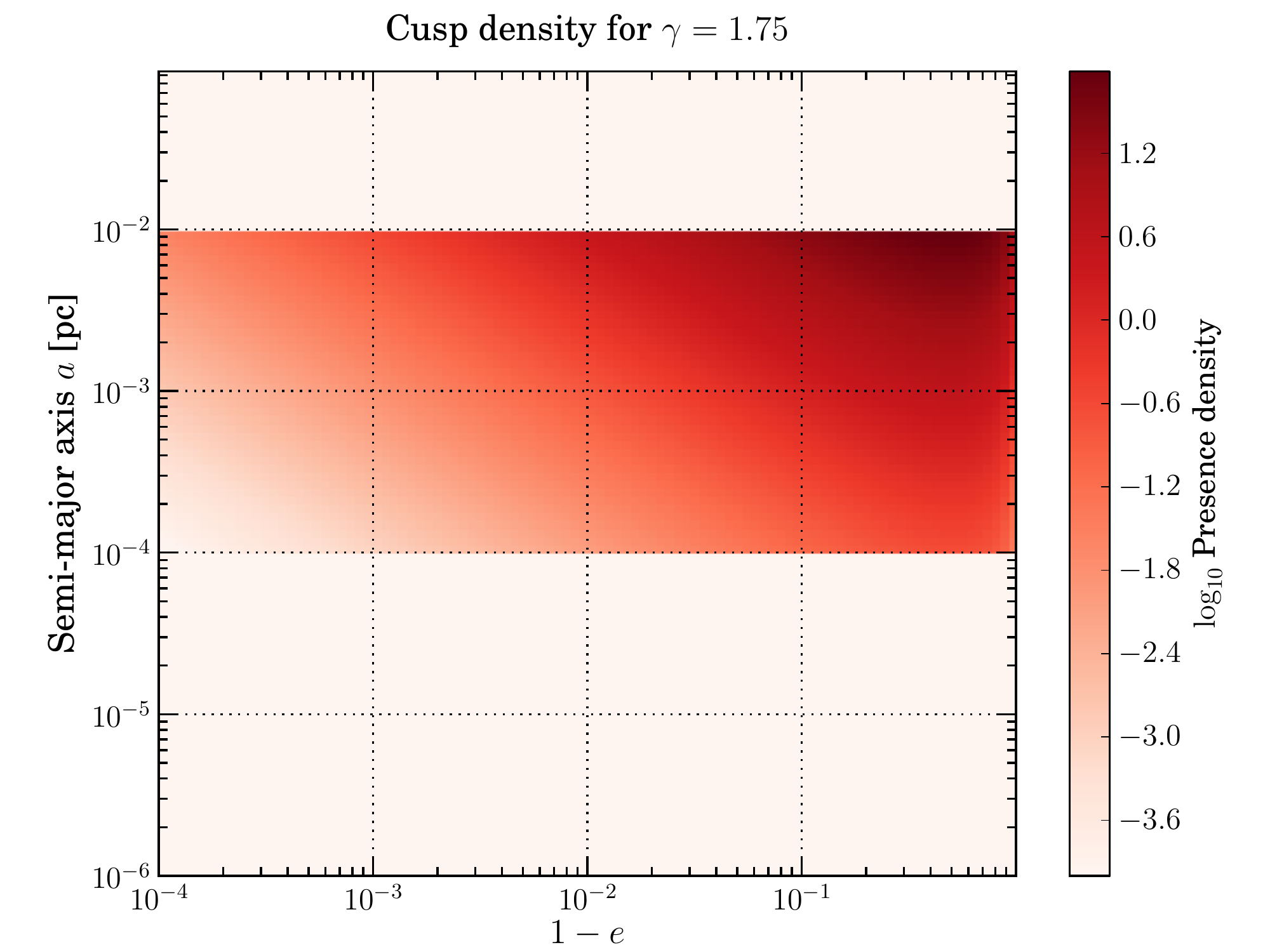}}
\caption
   {  
Theoretical distribution for the density of presence 
of a cusp of power-law 1.75 around a MBH, using a truncated distribution.   
   }
\label{fig.cusp2d}
\end{figure}

\begin{figure*}
\resizebox{\hsize}{!}
          {\includegraphics[scale=1,clip]{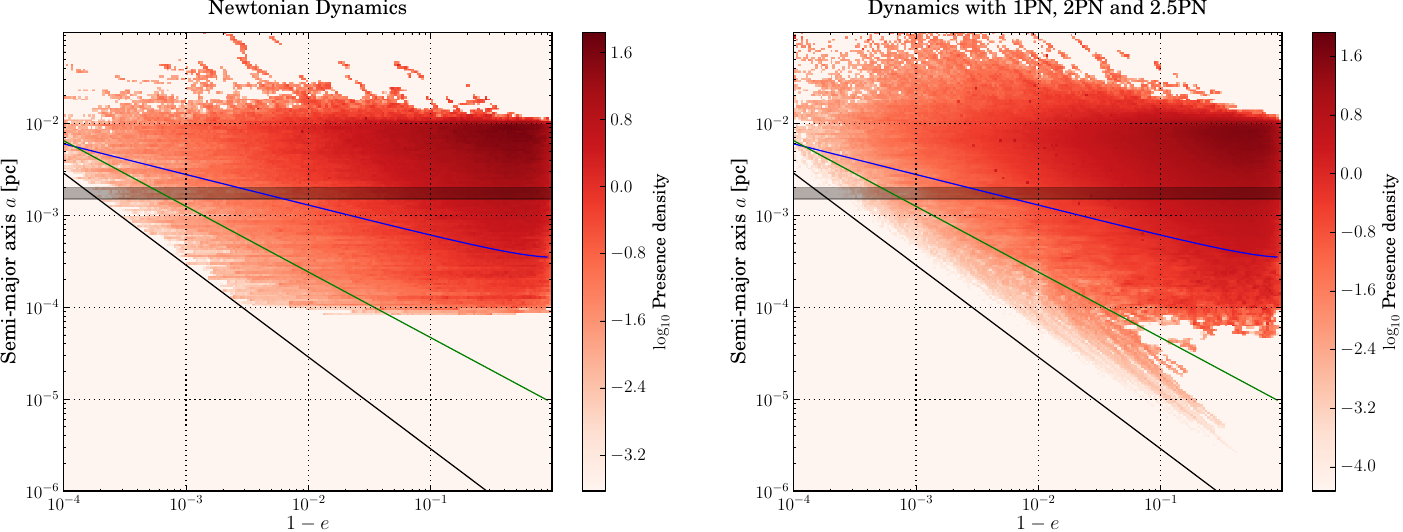}}
\caption
   {  
Integrated presence density for the Newtonian (left panel) and the
relativistic case (right panel). The shaded box marks the region of
the slice analyzed in Fig. \ref{fig.dndx}. The lines indicate the position of
the Schwarzschild barrier with $C_{\rm SB} = 0.35$ ({\itshape blue}) and the
limit for capture onto inspiral orbits for non-resonant relaxation ({\itshape
green}).   
   }
\label{fig.DensityPresence}
\end{figure*}

In order to quantify the nature of the ``Schwarzschild barrier'', we first plot
the normalized presence density as a histogram in the $(a, 1-e)$ plane for the
Newtonian case, Fig. \ref{fig.DensityPresence} (left panel) and the
relativistic case (right panel), and we give the theoretical distribution in
figure \ref{fig.cusp2d}. If we consider our specific setup, there are 3
different regions in the $(a,1-e)$ plane where different mechanisms are
efficient. In the right-most region, where pericenters are large, RR plays the
dominant role. The left border of this region is roughly given by the
appearance of the Schwarzschild precession which inhibits the BHs from
experiencing coherent torques \citep{BremAmaroSeoaneSopuerta2012}. We
derive that $C_{\rm SB} = 0.35$.

\section{Only fake plunges survive: Real EMRIs}

For a fast spinning MBH, the separatrix for prograde orbits is shifted to
significantly lower $a$ values, with a corresponding higher value of the
critical semi-major axis, corresponding to the point PP in
figure~\ref{fig.Timescales}.  As we have explained above, it is this effect
which can lead to a significant increase in the EMRI rate, combined with the
fact that the critical point for retrograde orbits (PR) is much less affected
and that an isotropic orbit distribution is expected, thanks to relaxational
processes.  However this increase in EMRI rate would can be thwarted by vector
RR if this process can change the orbital orientation of a SBH after it has
crossed the ``$t^{}_{\rm GW}=t^{}_{\rm r,\,peri}$'' line and before it has completed
its GW-driven inspiral, i.e.\ on a timescale shorter than $t^{}_{\rm GW}$. Indeed,
if the orbit becomes significantly less prograde as the the inspiral takes
place, due to RR, the separatrix moves up and the SBH might suddenly find
itself on a plunge orbit.

\begin{figure}
{\includegraphics[width=\columnwidth]{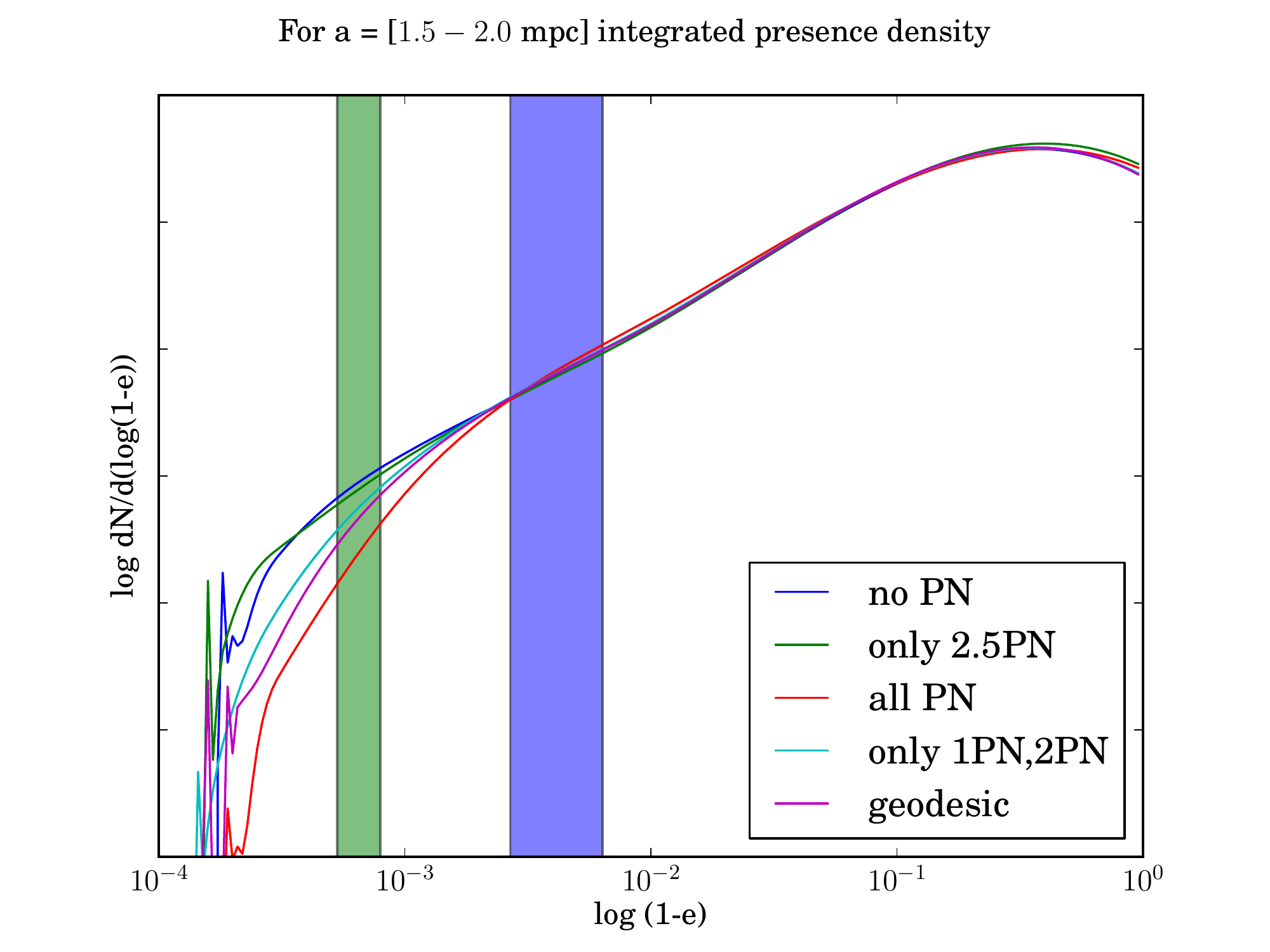}}
\caption
   {Integrated presence density for a semi-major axis slice $1.5$ mpc $<$ a $<$ $2.0$ mpc. The shaded boxes represent the position of the Schwarzschild barrier with $C_{\rm SB} = 0.35$ ({\itshape blue}) and the limit for capture on to inspiral orbits for non-resonant relaxation ({\itshape green}).
   }
\label{fig.dndx}
\end{figure}

To check for this possibility, we also plot, in figure~\ref{fig.Timescales},
a long-dashed line corresponding to the condition $t^{}_{\rm GW}=t^{}_{\rm RR,\,v}$,
with $t^{}_{\rm GW}<t^{}_{\rm RR,\,v}$ on the left of this line. SBHs that cross the
``$t^{}_{\rm GW}=t^{}_{\rm r,\,peri}$'' line while on the left side of the
``$t^{}_{\rm GW}=t^{}_{\rm RR,\,v}$'' line keep their orbital orientation during their inspiral
and complete it without abrupt plunge. One can see that, for our choice of
parameters, this is the case for all prograde orbits. On the other hand,
retrograde orbits can cross the ``$t^{}_{\rm GW}=t^{}_{\rm r,\,peri}$'' line while RR
is still effective enough to change their orientation during inspiral. However,
the effect of RR on retrograde orbits cannot reduce significantly the total
EMRI rate and may even increase it slightly because (1) these orbits contribute
less that the prograde ones (and more to the plunge rate) and (2)
statistically, RR is more likely to make the orbit become less retrograde which
pushes down the separatrix.

\begin{figure*}
\resizebox{\hsize}{!}
          {\includegraphics[scale=1,clip]{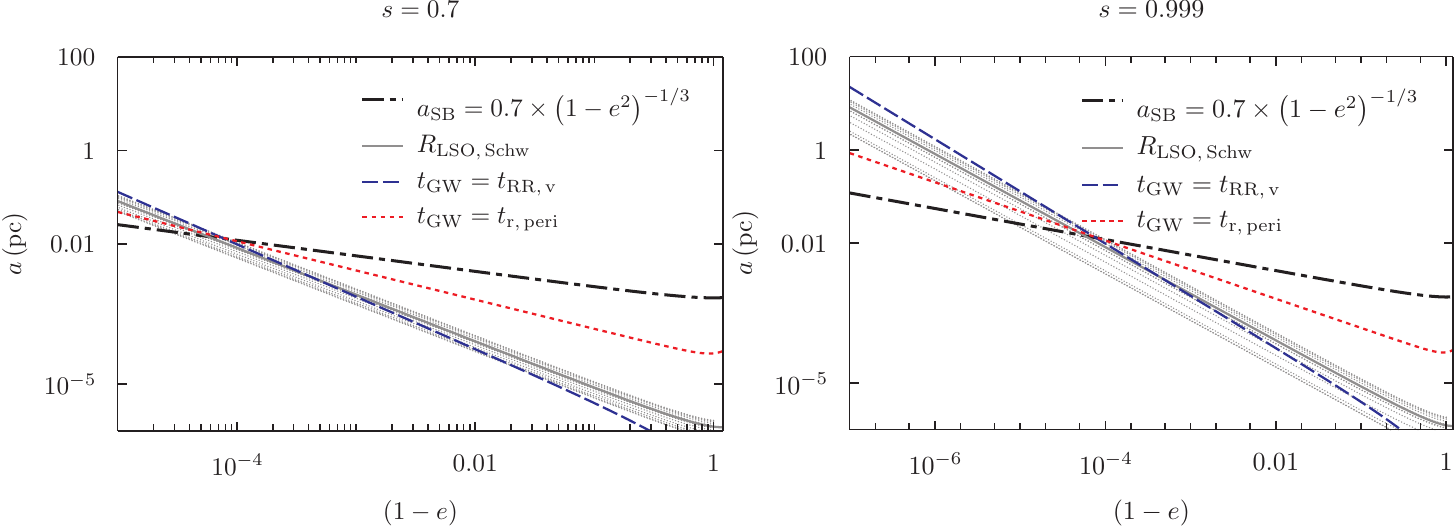}}
\caption
   {  
{\em Left panel:}
Relation between different timescales in the $s=0.7$ case.
We display the Schwarzschild separatrix as a solid, black line and the separatrices
for different inclinations with different curves in light grey. The dashed, blue line
shows the value of $a$ and $1-e$ for which the vectorial resonant relaxation timescale
($t^{}_{\rm RR,\,v}$) is equal to the gravitational loss timescale ($t^{}_{\rm GW}$). The
dashed, dotted line corresponds to the values of $a$ and $1-e$ for which the relaxation
time at periapsis ($t^{}_{\rm r,\,peri}$) equals the gravitational loss timescale.
{The dashed-dotted, black curve describes the ``Schwarzschild barrier''.}
{\em Right panel: }
Same as the left panel but for a spin value of $s=0.999$.
   }
\label{fig.Timescales}
\end{figure*}

Moreover, as we can see in figure \ref{fig.Timescales}, 
for semi-major axis with values approximately $a
\gtrapprox 0.03$ pc, the barrier lies well below the last separatrix, and so
EMRIs originating from these area, the ``plunges'' we have discussed in this
work, will not suffer this quenching in phase space.

\section{Conclusions}

The event rate of ``plunges'' is much larger than that of EMRIs, as a number of
different studies by different authors using different methods find.  Up to now
spin effects of the central MBH have been always ignored. Hence, the question
arises, whether a plunge is really a plunge when the central MBH is spinning.
This consideration has been so far always ignored.

So as to estimate EMRI event rates, one needs to know whether the orbital
configuration of the compact object is stable or not, because this is the
kernel of the difference between an EMRI and a plunge.  In this paper we take
into account the fact that the spin makes the LSO to be much closer to the
horizon in the case of prograde orbits but it pushes it away for retrograde
orbits.  Since the modifications introduced by the spin are not symmetric with
respect to the non-spinning case, and they are more dramatic for prograde
orbits, we prove that the inclusion of spin increases the number of EMRI events
by a significant factor. The exact factor of this enhancement depends on the
spin, but the effect is already quite important for spins around $s \sim 0.7$.

We also prove that these fake plunges, ``our'' EMRIs, do spend enough cycles
inside the band of eLISA to be detectable, i.e. they are to be envisaged as
typical EMRIs.  We note here that whilst it is true that EMRIs very near the
new separatrix shifted by the spin effect will probably contribute not enough
cycles to be detected, it is equally true for the old separatrix
(Schwarzschild, without spin).  In this sense, we find that the spin increases
generically the number of cycles inside the band for prograde EMRIs in such a
way that EMRIs very near to the non-spin separatrix, which contributed few
cycles, become detectable EMRIs.  In summary, spin increases the area, in
configuration space of detectable EMRIs.  We predict thus that EMRIs will be
highly dominated by prograde orbits.

We also demonstrate that these new kind of EMRIs we describe here originate in
a region of phase-space such that they will be ignorant of the Schwarschild
barrier. The reason for that is that they are driven by two-body relaxation and
not resonant relaxation.  While the boost in EMRI rates due to resonant
relaxation is affected by the Schwarzschild barrier, so that ``standard'' EMRIs
run into the problem of having to find a way to cross it, our ``plunge-EMRIs''
are already in the right place and led by two-body relaxation. The barrier
affects the productions of EMRIs via torques, but not two-body relaxation,
which is the mechanism producing the ``plunge-EMRIs''.  Moreover, because the
supression in the rates is severe for those EMRIs with semi-major axis with
values approximately $a \gtrapprox 0.03$ pc, we predict that the rates will be
dominated by the kind of EMRIs we have described in this work.

\section*{Acknowledgments}

PAS is indebted with S. Komossa and R. Saxton for asking him to give an invited
talk at the The Tidal Disruption Workshop.
This work has been supported by the Transregio 7
``Gravitational Wave Astronomy'' financed by the Deutsche
Forschungsgemeinschaft DFG (German Research Foundation).
CFS acknowledges support from the Ram\'on y Cajal Programme of the Spanish Ministry of Education 
and Science, contract 2009-SGR-935 of AGAUR,
and contracts FIS2008-06078-C03-03, AYA-2010-15709, and FIS2011-30145-C03-03 of MICCIN. 
We acknowledge the computational resources provided by the BSC-CNS
(AECT-2011-3-0007) and CESGA (contracts CESGA-ICTS-200 and CESGA-ICTS-221).

\end{document}